\documentclass[11pt]{article}
\usepackage[utf8]{inputenc}
\usepackage{amsmath}
\usepackage{amsfonts}
\usepackage{amssymb}
\usepackage{graphicx}
\usepackage{siunitx}
\usepackage{xcolor}
\usepackage{cite}
\usepackage{tikz}
\usepackage{bm,dsfont} 
\usepackage{float}
\usepackage{pgfplots}
\usepgfplotslibrary{fillbetween}
\usepackage{url}
\usepackage{siunitx}
\usepackage{hyperref}
\usepackage{xcolor}
\usepackage{array,booktabs}
\usepackage{tikz-cd}
\hypersetup{linktocpage,pdfauthor={Omar},colorlinks=true,urlcolor=purple,linkcolor=purple,citecolor=purple}

\tikzset{
	> = LaTeX,
	pics/lisa/.style = {
		code = {
			\draw [densely dashed, thick] (0, 1) -- (210:1) -- (330:1) -- cycle;
			\foreach \th in {90, 210, 330} {
				\draw [thick, fill = white] (\th:1) circle [radius = 0.2];
			}
		}
	}
}

\definecolor{lime}{HTML}{A6CE39}
\DeclareRobustCommand{\orcidicon}{
	\begin{tikzpicture}
	\draw[lime, fill=lime] (0,0) 
	circle [radius=0.16] 
	node[white] {{\fontfamily{qag}\selectfont \tiny ID}};
	\draw[white, fill=white] (-0.0625,0.095) 
	circle [radius=0.007];
	\end{tikzpicture}
	\hspace{-2mm}
}

\foreach \x in {A, ..., Z}{%
	\expandafter\xdef\csname orcid\x\endcsname{\noexpand\href{https://orcid.org/\csname orcidauthor\x\endcsname}{\noexpand\orcidicon}}
}

\newcommand{\comment}[1]{}

\DeclareMathAlphabet{\mathfs}{U}{rsfs}{m}{n}                     %

\newcommand{\n}{\nonumber}
\newcommand{\be}{\nopagebreak[3]\begin{equation}}
	\newcommand{\ee}{\end{equation}}
\newcommand{\bee}{\nopagebreak[3]\begin{equation*}}
	\newcommand{\eee}{\end{equation*}}
\newcommand{\ba}{\nopagebreak[3]\begin{eqnarray}}
	\newcommand{\ea}{\end{eqnarray}}
\newcommand{\baa}{\nopagebreak[3]\begin{eqnarray*}}
	\newcommand{\eaa}{\end{eqnarray*}}
\newcommand{\bal}{\nopagebreak[3]\begin{aligned}}
	\newcommand{\eal}{\end{aligned}}

\newcommand{\bseq}{\nopagebreak[3]\begin{subequations}}
	\newcommand{\eseq}{\end{subequations}\noindent}

\numberwithin{equation}{section}

\begin{document}
\title{ \textbf{AdS Carroll Structures from Poincaré Isomorphism: ~~~~~~~~~~
		Asymptotic Symmetry Analysis}}
\author{Luis Avilés$^{1,2}$\orcidE{},
Joaquim Gomis$^{3}$\orcidF{},
Diego Hidalgo$^{4}$\orcidD{},
Omar Valdivia$^{1,2}$\orcidA{}.
\bigskip \\
{\small \textit{$^{1}$ Facultad de Ciencias, Universidad Arturo Prat, Avenida Arturo Prat Chac\'on 2120, Iquique, Chile.}} \\
{\small \textit{$^{2}$ Instituto de Ciencias Exactas y Naturales (ICEN), Universidad Arturo Prat,  Playa Brava 3256, Iquique, Chile.}}\\
{\small \textit{$^{3}$ Departament de F\'isica Qu\`{a}ntica i Astrof\'isica and Institut de Ci\`{e}ncies del Cosmos (ICCUB),}}\\{ \small  \textit{  Universitat de Barcelona, Mart\'i i Franqu\`{e}s 1, E-08028 Barcelona, Spain.}}\\
{\small \textit{$^{4}$ Science Institute, University of Iceland, Dunhaga 3, 107 Reykjav\'ik, Iceland.%
}}}
\maketitle
	\flushbottom
	
	\abstract{\noindent Starting from the isomorphism between the AdS Carroll and Poincaré algebras, we map the three-dimensional asymptotically flat solutions of Poincaré gravity into an AdS Carroll spacetime. We show the mapped solutions satisfy the field equations of the Chern-Simons formulation of AdS Carroll gravity and exhibit a Carroll geometry structure. Despite the presence of a negative cosmological constant in the mapped spacetime, the algebra of the canonical generators of the asymptotic symmetries is given by the $\mathfrak{bms}_3$ algebra.} 
	\newpage
	
	\section{Introduction}
	Einstein's gravity in \((2+1)\)-dimensional spacetime provides a comprehensive framework for exploring the features of higher-dimensional gravity and the quantum nature of gravitational theories \cite{Achucarro:1986uwr, Witten:1998qj}. Despite the absence of local degrees of freedom, the theory manifests significant global properties \cite{Deser:1983nh}. In the limit of a vanishing cosmological constant, the symmetries of this gravitational theory are dictated by the Poincaré algebra $\mathfrak{iso}(2,1)$.
	The extension of this symmetry has played an essential role in many relativistic systems, such as its enhancement to include conformal transformations \cite{Polchinski:1987dy, Qualls:2015qjb}, its extension to include fermionic generators known as super-Poincaré \cite{Fuentealba:2015jma, Fuentealba:2015wza}, and the extension to the \(\mathfrak{bms}\) algebra, which emerges from the structure of \((3+1)\)-dimensional asymptotically flat spacetimes at null infinity \cite{Sachs:1962zza, Bondi:1962px}. 
	In the case of $(2+1)$-spacetime dimensions, this analysis has been studied in \cite{Barnich:2006av}.\\
	When a negative cosmological constant is included, isometries are described by the anti-de Sitter (AdS) group \(SO(2,2)\). For \((2+1)\)-dimensional AdS spacetimes (AdS\(_3\)), there is a comprehensive understanding of the asymptotic symmetries, global charges, central extensions \cite{Brown:1986nw}, space of solutions \cite{Banados:1992wn}, and their conformal field theory interpretation \cite{Strominger:1997eq}. Notably, one of the most significant discoveries is the Bañados-Teitelboim-Zanelli (BTZ) black hole solution, which exhibits thermodynamic properties such as temperature and entropy \cite{Banados:1992wn, Banados:1992gq}.\\
	In an effort to explore flat spacetime holography and to model more realistic scenarios of the AdS/CFT correspondence \cite{Maldacena:1997re}, the case of \((2+1)\)-dimensional flat gravity has been considered \cite{Susskind:1998vk, Polchinski:1999ry, Arcioni:2003td, Mann:2005yr, Arcioni:2003xx, Giddings:1999jq}. Unlike AdS\(_3\) gravity, \((2+1)\)-dimensional Poincaré gravity does not contain a black hole solution. However, there exists a notion of a Cauchy horizon associated with a cosmological solution that exhibits thermodynamic properties within the grand canonical ensemble \cite{Barnich:2012xq}. This cosmological solution and its thermodynamics are interpreted as the flat limit of one of the inner BTZ horizons \cite{Castro:2012av, Detournay:2012ug}. Furthermore, there are similar solutions for gravity theories based on modifications of the Poincaré algebra, such as the Maxwell algebra \cite{Concha:2018zeb}.
	
	
	Other explorations of the Poincaré symmetry algebra include its possible contractions. There exist two main \.{I}nönü-Wigner contractions for the Poincaré algebra: the so-called Galilei and Carroll algebras. These two contractions describe many interesting and useful non-Lorentzian theories (see, e.g., reviews \cite{Bergshoeff:2022eog, Oling:2022fft, Hartong:2022lsy}). To carry out these contractions, one considers a dimensionless constant \(\omega\). The Galilei algebra is obtained in the \(\omega \to \infty\) limit, and the Carroll algebra in the \(\omega \to 0\) limit. In reference \cite{Bacry:1968zf}, Bacry and Lévy-Leblond classified all possible kinematical algebras that include spacetime translations, spatial rotations, and boosts (see also \cite{Figueroa-OFarrill:2017ycu} for a new approach to this classification). In particular, contractions of the AdS algebra lead to generalizations of the Galilei and Carroll algebras, including a cosmological constant. The extension of the Carroll algebra to include a negative cosmological constant turned out to be isomorphic to the Poincaré algebra, which does not depend on the dimension of spacetime. The authors referred to this extension of the Carroll algebra as the "Para-Poincaré" algebra. Here, we use modern terminology and refer to it as the AdS Carroll algebra.

	In light of the aforementioned isomorphism, we explore the mapping of the flat cosmological solution into Carroll geometries using the first-order formulation of \((2+1)\)-dimensional Einstein gravity and the isomorphism between Poincaré and AdS Carroll algebras. We demonstrate that the AdS Carroll gravity theory can be derived by gauging the AdS Carroll algebra or by directly applying the isomorphism.
	By examining the asymptotic conditions of flat cosmology, which are governed by the \(\mathfrak{bms}_3\) algebra, it becomes evident that AdS Carroll gravity shares the same infinite-dimensional algebra. The absence of the cosmological constant in this asymptotic structure is due to the isomorphism and reflects the field theory counterpart of the AdS Carroll particle \cite{Casalbuoni:2023bbh}. In other words, the asymptotic structure of AdS Carroll gravity in any spacetime dimension is always described by that of Poincaré gravity theory.
	Despite the algebraic structure underpinning this isomorphism, the solution mapped into a Carroll structure possesses a different physical interpretation because the generators act differently on the corresponding homogeneous spacetimes. This work aims to extend the worldline approach of the relativistic particle and the particle in flat AdS Carroll spacetime \cite{Bergshoeff:2022eog}, where the aforementioned isomorphism results in an explicit interchange of the corresponding dynamical variables.

	The remainder of the paper is organized as follows. In Section \ref{Section:isomorphism}, we establish the notation and describe the isomorphism between the AdS Carroll and Poincaré algebras in \((2+1)\)-dimensions. After detailing the isomorphism between the algebra generators, we extend it to the associated gauge fields of both algebras and present the AdS Carroll gravity solution in the first-order formalism in Section \ref{Section:AdSCarrollsolution}. We also derive the asymptotic conditions for the AdS Carroll gravity theory gauge fields, which generate the \(\mathfrak{bms}_3\) algebra as an asymptotic symmetry. In Section \ref{Geometrysection}, we construct the Carroll structure associated with this solution and provide its physical descriptions. In Section \ref{Section:discussion}, we summarize our results and discuss future research directions.

	\section{Isomorphism between AdS Carroll and Poincaré algebra}\label{Section:isomorphism}

	This section explores the features of the isomorphism between the Poincaré and AdS Carroll algebras, which result in a distinct physical interpretation of the corresponding homogeneous spaces. This isomorphism is based on a straightforward exchange of roles between boosts and spatial momenta in the two algebras. In the field theory context, this leads to a mapping between solutions in \((2+1)\)-dimensional Poincaré and AdS Carroll gravity theories and their asymptotic structures, among other implications.

	\subsection{The isomorphism}

	The AdS\(_3\) algebra \(\mathfrak{so}(2,2)\) admits three types of \.{I}nönü-Wigner contractions. The first, known as the flat limit, results in the Poincaré algebra \(\mathfrak{iso}(2,1)\). The second, called the non-relativistic limit, yields the Newton-Hooke (NH\(_-\)) algebra. The third, which involves taking the speed of light to zero, produces the AdS Carroll algebra. In this note, we will focus on the latter.
	
	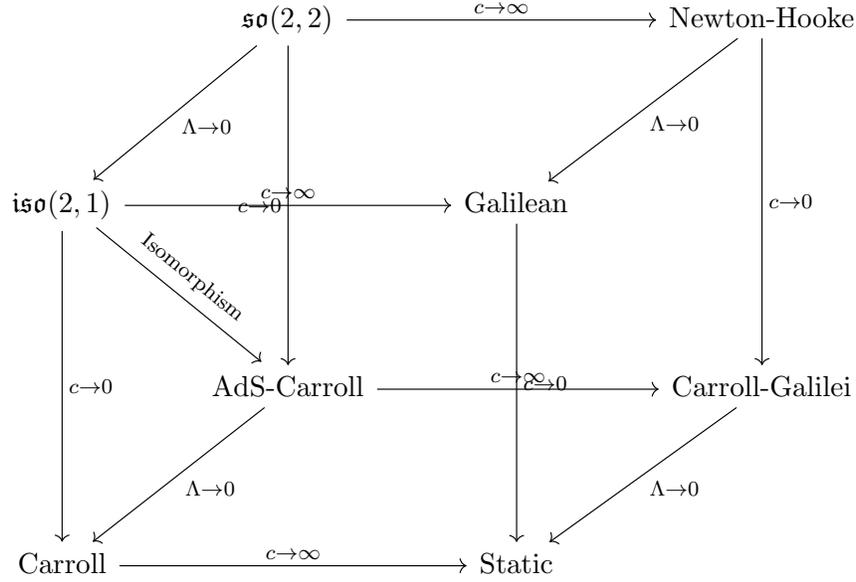
\begin{figure}[h]
		\centering
		\begin{tikzcd}[column sep=1cm,row sep=1.8cm]
			& |[alias=A]| \mathfrak{so}(2,2) \arrow[dl,"\Lambda \rightarrow 0"] \arrow[dd,"c\rightarrow 0"'] \arrow[rr,"c\rightarrow \infty"] & & |[alias=B]| \text{Newton-Hooke} \arrow[dl,"\Lambda \rightarrow 0"] \arrow[dd, "c \rightarrow 0"] \\
			|[alias=C]| \mathfrak{iso}(2,1) \arrow[dd,"c \rightarrow 0"] \arrow[rr,"c \rightarrow \infty"] \arrow[from=C, to=E, shift left=.35ex,"\text{Isomorphism}"{sloped, yshift = 2pt, xshift=0pt} ]& & |[alias=D]| \text{Galilean} \arrow[dd,"c \rightarrow 0"] \\
			& |[alias=E]| \text{AdS-Carroll} \arrow[dl,"\Lambda \rightarrow 0"] \arrow[rr,"c \rightarrow \infty"] & & |[alias=F]| \text{Carroll-Galilei} \arrow[dl,"\Lambda \rightarrow 0"] \\
			|[alias=G]| \text{Carroll} \arrow[rr,"c \rightarrow \infty"] & & 	|[alias=H]| \text{Static}
		\end{tikzcd}
		\caption{\scriptsize Diagram illustrating the relationships between various kinematical algebras through limiting processes. The arrows represent different limits: \(\Lambda \rightarrow 0\) denotes the zero cosmological constant limit, while \(c \rightarrow 0\) and \(c \rightarrow \infty\) indicate the speed of light approaching zero or infinity, respectively. An isomorphism between \(\mathfrak{iso}(2,1)\) and AdS-Carroll algebra is highlighted, showing the connections between these algebras and their associated kinematical groups.}
	\end{figure}

	We begin by considering the \((2+1)\)-dimensional Poincaré algebra, which is spanned by the dual Lorentz generators \(\tilde{J}_A\). These generators are related to the Lorentz generators by the relations
	
	\begin{equation}
		\tilde{J}_A = - \frac{1}{2} \epsilon_{ABC} \tilde{J}^{BC} \quad \text{and} \quad \tilde{J}_{AB} = \epsilon_{ABC} \tilde{J}^C,\label{eq:Poincarealgebranormal}
	\end{equation}
	where \(A\) ranges over the indices \((0,a)\) with \(a = 1, 2\). The algebra also includes the momentum generators \(\tilde{P}_A\). These generators satisfy the commutation relations
	\begin{equation}
		\left[\tilde{J}_A, \tilde{J}_B\right] = \epsilon_{ABC} \tilde{J}^C, \quad \left[\tilde{J}_A, \tilde{P}_B\right] = \epsilon_{ABC} \tilde{P}^C,
	\end{equation}
	where the indices are raised and lowered using the diagonal Minkowski metric \(\eta_{AB} = \text{diag}(-1,1,1)\), and the Levi-Civita symbol satisfies \(\epsilon_{012} = 1 = -\epsilon^{012}\).
	The non-degenerate and invariant bilinear form associated with these commutation relations is given by
	\begin{equation}
		\langle \tilde{J}_A, \tilde{J}_B \rangle = - \alpha_0 \, \eta_{AB}, \quad \langle \tilde{J}_A, \tilde{P}_B \rangle = - \alpha_1 \, \eta_{AB},
	\end{equation}
	where \(\alpha_0\) and \(\alpha_1\) are arbitrary constants. It is important to note that \(\alpha_1 \neq 0\) to ensure that the bilinear form is non-degenerate.
	

	In the classification of "kinematical groups" carried out by Bacry and Lévy-Leblond \cite{Bacry:1968zf}, it was observed that an isomorphism exists between \(\mathfrak{iso}(3,1)\) and the AdS Carroll algebra. The later is generated by time translations \(H\), spatial translations \(P_a\), spatial rotations \(J\), and Carroll boosts \(B_a\), which satisfy the following non-zero commutation relations
	
	\begin{subequations}\label{Eq:AdSCarrollalgebra}
		\begin{align}
			\left[B_{a},P_{b}\right] &= \delta_{ab}\, H, \quad &\left[P_{a},P_b\right] &= -\frac{1}{\ell^2}\,\epsilon_{ab}\, J,\\ 
			\left[J,P_a\right] &= \epsilon_{ab}\, P_{b}\,, \quad &\left[ P_a,H \right] &= -\frac{1}{\ell^2}\, B_a\,,\\
			\left[J,B_{a}\right] &= \epsilon_{ab} \,B_b\,,
		\end{align}
	\end{subequations}
	where \(\ell > 0\) denotes the AdS Carroll radius, which is related to the cosmological constant by \(\Lambda = -1/\ell^2\). Here, \(\delta_{ab}\) refers to the Kronecker delta, and the two-dimensional Levi-Civita symbol is defined such that \(\epsilon_{12} = \epsilon^{12} = 1\). The flat limit, \(\ell \to \infty\), results in the three-dimensional Carroll algebra. Explicitly, the isomorphism in \((2+1)\)-dimensions is given by \cite{Bergshoeff:2022eog}
	
	\begin{equation}\label{eq:mapgenerators}
		\tilde{J} = J, \quad \tilde{J}_a = -\ell \, \epsilon_{ab}\, P_b, \quad \tilde{H} = H, \quad \tilde{P}_a = -\frac{1}{\ell}\, B_a\,,
	\end{equation}
	with \(\tilde{J} \equiv \tilde{J}_0\) and \(\tilde{H} \equiv \tilde{P}_0\).
	
	It is essential to emphasize that the physical interpretations of these systems are fundamentally distinct. As analyzed in \cite{Bergshoeff:2022eog}, for particle actions, the same kinematical Lie group—in this case, the Poincaré group—can correspond to inequivalent homogeneous spacetimes. Specifically, what constitutes a translation for a massive spinless particle in Poincaré spacetime is analogous to a Carroll boost for a Carroll particle in AdS Carroll spacetime. By extending this isomorphism to field theory, we will demonstrate that it provides an alternative framework for constructing \((2+1)\)-dimensional AdS Carroll geometries in gravity theories.

	\section{AdS Carroll gravity solution from the $(2+1)$-Poincaré cosmology}\label{Section:AdSCarrollsolution}

	Three-dimensional Einstein gravity theory admits an asymptotically flat cosmological solution exhibiting thermodynamic properties \cite{Barnich:2012xq}. This solution can be interpreted as the flat-space limit of the BTZ black hole. In this section, we aim to map this solution into a Carroll geometry within the context of the Chern-Simons formulation of gravity. We will first outline the mapping between the gauge fields and subsequently derive the asymptotic structure of the resulting AdS Carroll spacetime.
	
	\subsection{Map between Gauge Fields}
	
	Our objective is to derive an AdS Carroll gravity solution with a non-trivial asymptotic structure. To achieve this, we need to elucidate the relationship between the gauge fields of both theories. We consider the one-form gauge connection \( A \), valued on the basis of the Poincaré and AdS Carroll algebras, given by
	
	\begin{equation}
		A = \tilde{e}\tilde{H} + \tilde{e}^a \tilde{P}_a + \tilde{\omega}\tilde{J} + \tilde{\omega}^a \tilde{J}_a = \tau H + e^a P_a + \omega J + \omega^a B_a\,,\label{derivinggaaugefields}
	\end{equation}
	where the one-form gauge fields on the right-hand side correspond to those of the AdS Carroll algebra. Utilizing the isomorphism \eqref{eq:mapgenerators}, we obtain the following relationships between the gauge fields of both algebras
	
	\begin{subequations}\label{mapCP}
		\begin{align}
			\tilde{e} &= \tau\,, \\
			\tilde{e}^a &= -\ell\, \omega^a\,, \\
			\tilde{\omega} &= \omega\,, \\
			\tilde{\omega}_a &= \frac{1}{\ell}\, \epsilon_{ab}\, e^b\,.
		\end{align}
	\end{subequations}
	We observe that this isomorphism results in an exchange of roles between the dreibeins and the spin connection within the one-form gauge field. This phenomenon generates a novel geometrical interpretation of the flat cosmology within the AdS Carroll gravity framework. It is noteworthy that this isomorphism can also be extended to higher dimensions, independently of the Chern-Simons formulation.

	\subsection{Chern-Simons formulation for AdS Carroll algebra}

	To analyze the dynamics and asymptotic structure of the AdS Carroll spacetime, we start by considering the Chern-Simons formulation of AdS Carroll symmetry in \((2+1)\)-dimensions \cite{Matulich:2019cdo, Gomis:2019nih}. First, we note that the algebra \eqref{Eq:AdSCarrollalgebra} is equipped with a non-degenerate and invariant bilinear form given by
	
	\begin{equation}
		\langle J\,,J \rangle = \beta_2\,, \quad \langle P_a,P_b \rangle = -\frac{\beta_2}{\ell^2}\, \delta_{ab}\,, \quad \langle J,H \rangle = \beta_1\,, \quad \langle B_a,P_b \rangle = -\beta_1\, \epsilon_{ab}\,,\label{Eq:AdSpairing}
	\end{equation}
	where \(\beta_1\) and \(\beta_2\) are arbitrary constants, with \(\beta_1 \neq 0\). Henceforth, we raise and lower indices with the spatial metric \(\delta_{ab}\) in AdS Carroll gravity to avoid introducing a minus sign. The Chern-Simons form for the gauge connection \(A\), constructed with the invariant bilinear form \eqref{Eq:AdSpairing}, defines the action for AdS Carroll symmetry as
	
	\begin{equation}
		S[A] = \frac{k}{4\pi}\, \int_{\mathcal{M}}\, \left\langle A \wedge \mathrm{d} A + \frac{1}{3} \, A \wedge [A, A] \right\rangle\,,\label{eq:CSformula}
	\end{equation}
	with \(k\) being the Chern-Simons level and \(\mathcal{M}\) a \((2+1)\)-dimensional Carroll manifold. In light of \eqref{Eq:AdSCarrollalgebra}, \eqref{derivinggaaugefields}, and \eqref{Eq:AdSpairing}, the action \eqref{eq:CSformula} becomes
	\begin{equation}
		S[A] = \frac{k}{2\pi}\,\int_{\mathcal{M}}\, \left( \epsilon_{ab}\, e^a \,R^b (B) - \tau R(J) + \frac{1}{\ell^2} \, \epsilon_{ab} e^a e^b \tau \right)\,,\label{eq:AdSCarrollaction}
	\end{equation}
	where we set \(\beta_2 = 0\) and \(\beta_1 = -1\) for simplicity. Here and in the following, the two-form curvatures \(F = \mathrm{d}A + \frac{1}{2}[A, A]\) are given by\footnote{Henceforth, the wedge product \(\wedge\) between differential forms is implied, i.e., \(\omega \omega^b = \omega \wedge \omega^b\).}
	\begin{subequations}\label{eq:AdsCarrollcurvatures}
		\begin{align}
			R(H) &= \mathrm{d}\tau + \omega_a e_a\,, \\
			R(J) &= \mathrm{d}\omega - \frac{1}{2 \ell^2} \, \epsilon_{ab} e_a e_b\,, \\
			R_a (P) &= \mathrm{d}e_a + \epsilon_{ab} e_b \omega\,, \\
			R_a (B) &= \mathrm{d}\omega_a + \frac{1}{\ell^2} \, \tau e_a - \epsilon_{ab} \omega \omega_b\,.
		\end{align}
	\end{subequations}
	
	It can be verified that the AdS Carroll two-form curvatures \(F = \mathrm{d}A + \frac{1}{2}[A, A]\) are related to the Poincaré curvatures as follows
	
	\begin{subequations}
		\begin{align}
			R(\tilde{H}) &= R(H)\,, \quad &R^a (\tilde{P}) = -\ell\, R^a(B)\,, \\
			R(\tilde{J}) &= R(J)\,, \quad &R_a (\tilde{J}) = \frac{1}{\ell} \,\epsilon_{ab}\, R^b(P)\,,
		\end{align}
	\end{subequations}
	where tildes denote the Poincaré gauge fields. It is straightforward to check that \eqref{eq:AdSmetric} is a solution for all the AdS Carroll curvatures on the right-hand side of the last equation. The Chern-Simons framework forms the basic setup for defining the boundary conditions for the AdS Carroll gravity action \eqref{eq:AdSCarrollaction}.

	\subsection{Asymptotic structure}
	%
	To describe the asymptotic structure of the AdS Carroll manifold $\mathcal M$, we implement the set of BMS-like coordinates $x^{\hat \mu} =(u,r,\phi)$ with the boundary parametrized by the retarded time coordinate $-\infty <u<\infty$, a radial direction $0\leq r <\infty$, and an angular coordinate $\phi\sim \phi +2\pi$. One of the more salient features in $(2+1)$-dimensions Chern-Simons gravity is that we can remove the radial coordinate of the gauge field via the gauge transformation \cite{Banados:1994tn,Bunster:2014mua}
	\be\label{eq:gaugetransformation}
	A(u,r,\phi) = b^{-1} (r) \Bigg(\mathrm{d}+ a(u,\phi)\Bigg) b(r)\,,
	\ee
	with
	\be\label{eq:smallconnection}
	a(u,\phi) = a_u (u,\phi)\, \mathrm{d} u+a_\phi (u,\phi) \, \mathrm{d}\phi \,.
	\ee
	and the group element $b$ being a permissible group element in the sense that does not change the value of the charges.\footnote{For a permissible gauge transformation we mean that the variation of the Chern-Simons charges $\delta Q\left[\lambda,a_\phi \right]$ must be unaltered after that gauge transformation. Namely, let $\delta Q\left[\bar \lambda,\bar{a}_\phi \right]$ be the variation of the charge obtained after a gauge transformation. The charge is invariant if the quantity $ \Delta \, Q \equiv \delta Q\left[\bar{\lambda},\bar{a}_\phi \right]-\delta Q\left[\lambda,a_\phi\right] $ is equal to zero.} This gauge connection $a$ allows gaining calculation advantage to solve the field equations $f=\mathrm{d}a+\frac{1}{2}[a,a]=0$ and consequently the field equation associated to $A$.  For the solution presented below, we choose the following group element
	\be\label{eq:groupelement}
	b(r)= \exp \left( - \frac{r}{\sqrt{2}} \left(H- \frac{1}{\ell} \, B_2 \right) \right)\,.
	\ee
	The general solution of $(2+1)$-Poincaré gravity \cite{Barnich:2012xq} has a metric in a BMS-like gauge given by
	\be\label{eq:solutioninBMSgauge}
	ds^{2}_{P} = \mathcal M \, \mathrm{d}u^2 -2 \mathrm{d}u \, \mathrm{d}r + \mathcal N \, \mathrm{d}\phi \, \mathrm{d}u + r^2 \, \mathrm{d} \phi^2 \,,
	\ee
	where $\mathcal M $ and $\mathcal N $ are given by
	\be
	\mathcal{M}=\mathcal{P}(\phi)\,, \hspace{1cm} \mathcal{N}(u,\phi) = u\,\mathcal{P'}(\phi)+\mathcal{J}(\phi)  \,,
	\ee 
	with the prime denoting derivative along the cyclic coordinate $\phi$. For zero modes this metric corresponds to a flat cosmology \cite{Barnich:2012xq, Barnich:2012aw}.
	
	In this BMS gauge, the boundary conditions of the Poincaré gravity theory are described by \eqref{eq:smallconnection} in its corresponding basis. To map them into the Carroll gravity, we need to map the Poincaré generators by rewriting the Poincaré algebra in a lightcone form (see Appendix \ref{section:appendixlightcone}). Denoting the lightcone Poincaré generators with a hat symbol, the isomorphism now becomes
	\bseq\label{eq:mappinglightcone}
	\ba
	\hat{P}_0 &=& \frac{1}{\sqrt{2}}\, \left( H - \frac{1}{\ell}\, B_2\right)\,, \hspace{1.3cm} \hat{J}_0 =  -\frac{1}{\sqrt{2}}\, \left( J+ \ell \, P_1 \right) \,, \\
	\hat{P}_1 &=&- \frac{1}{\sqrt{2}}\, \left( H + \frac{1}{\ell}\, B_2\right)\,, \hspace{1cm} \hat{J}_1 =\frac{1}{\sqrt{2}}\, \left( J- \ell \, P_1 \right) \,,\\
	\hat{P}_2 &=&-\frac{1}{\ell}\, B_1\,, \hspace{3.0cm} \hat{J}_2 =\ell \, P_2\,.
	\ea
	\eseq
	From the relation \eqref{derivinggaaugefields} and using \eqref{eq:mappinglightcone}, the one-form gauge fields are now related as
	\bseq\label{eq:mapfieldsinlightcone}
	\ba
	\tau &= & \frac{1}{\sqrt{2}}\, \left(\tilde e^0- \tilde e^1 \right)\,, \hspace{2.0cm} \omega  =  \frac{1}{\sqrt{2}} \left( \tilde \omega^1 -\tilde \omega^0\right)\,, \\
	e^1 &= & -\frac{\ell}{\sqrt{2}}\,\left( \tilde \omega^0 +\tilde \omega^1 \right)\,, \hspace{1.4cm} \omega^1 = -\frac{1}{\ell}  \tilde e^2 \,,\\
	e^2 &= & \ell \, \tilde \omega^2\,, \hspace{3.5cm} \omega^2  =-  \frac{1}{\sqrt{2}\, \ell} \, \left(\tilde e^0+ \tilde e^1 \right)\,.
	\ea
	\eseq
	Then, in virtue of \eqref{eq:mappinglightcone} and \eqref{eq:mapfieldsinlightcone}, the boundary conditions \eqref{eq:smallconnection} for AdS Carroll spacetimes in $(2+1)$-dimensions become
	\bseq\label{eq:bbccAdSCarroll}
	\ba
	a_u & =& \frac{1}{\sqrt{2}} \left(  \frac{\mathcal{P}}{2} -1\right)  \, H -  \frac{1}{\sqrt{2}\, \ell} \, \left( \frac{\mathcal{P}}{2}  +1 \right)\, B_2\,,\\
	\nonumber  a_\phi &=&   \left( \frac{1}{2\sqrt{2}} \left( u\mathcal{P}^\prime + \mathcal J \right) \right) \, H  -\frac{\ell}{\sqrt{2}}\, \left(\frac{\mathcal{P}}{2} +1\right) \, P_1 +  \frac{1}{\sqrt{2}} \left(1-\frac{\mathcal{P}}{2}  \right) \, J  -    \left( \frac{1}{2\sqrt{2}\, \ell}\left( u\mathcal{P}^\prime + \mathcal J \right)  \, \right)\, B_2 \,. \\
	\label{eq:AdSCarrollbdrtyconditions}
	\ea
	\eseq
	One can check that the set of boundary conditions \eqref{eq:bbccAdSCarroll} lead to a well-defined variational principle for the action \eqref{eq:CSformula}. This can be checked by considering the general variation of the Chern-Simons action \eqref{eq:CSformula} as follows
	\be
	\delta S[A] = \frac{k}{2\pi} \int_{\mathcal M} \langle \delta A \,, F\rangle + \frac{k}{4\pi} \int_{\partial \mathcal M} \langle \delta A \,, A\rangle\,.   
	\ee
	The first term vanishes for $F=0$, while for the second term, one can check the boundary conditions \eqref{eq:AdSCarrollbdrtyconditions} kills it. Then, those boundary conditions lead to a consistent variational principle. 
	The gauge fields that one can extract from \eqref{eq:bbccAdSCarroll} via $g_{\hat{\mu} \hat{\nu}} = \hat{\eta}_{AB}\,{\tilde{e}^{A}}_{\hat{\mu}}\, {\tilde{e}^{B}}_{\hat{\nu}}$, with $\hat{\eta}_{AB}$ being the off-diagonal metric in Appendix \ref{section:appendixlightcone}, will
	allow us to construct the Carroll geometry for this solution. We will back to this point in Section \ref{Geometrysection}.

	\subsection{Gauge transformations preserving the boundary conditions}
	Now, we aim to determine the gauge transformations $\delta_\lambda a = \mathrm{d} \lambda + [a\,, \lambda]$ that preserve the boundary conditions \eqref{eq:bbccAdSCarroll}. Using the terminology of \cite{Benguria:1976in}, these gauge transformations split into two groups, proper and improper. The latter corresponds to those that change the state of the theory, and therefore they are associated with non-trivial canonical charges. Then, the gauge transformations that will be derived below correspond to those that act non-trivially modulo the proper gauge transformations. We consider the zero-form gauge parameter $\Xi$ as
	\be\label{eq:gaugeparameterwithoutr}
	\Xi (u,r,\phi) = b^{-1}(r) \, \lambda (u,\phi)\, b(r)\,, \hspace{0.5cm} \lambda(u,\phi) = \varepsilon H + \varepsilon^a P_a + \chi^a B_a +\chi J\,,
	\ee   
	where the functions $\varepsilon, \varepsilon^{a}, \chi^a$, and $\chi$ are functions of $u$ and $\phi$. With this expression, the equation $\delta_\lambda a = \mathrm{d} \lambda + [a,\lambda]$ gives us the following solutions for the gauge parameters
	\ba\label{eq:gaugetransf}
	\nonumber \varepsilon &= & \frac{1}{\sqrt{2}} \, \left(\frac{2\pi}{k}\left(Y \mathcal{J}+T\mathcal{P}\right)-T-T''\right) +  \frac{1}{\sqrt{2}}\, \left( Y^\prime\left( \frac{2\pi}{k} \mathcal P - 1  \right) +\frac{2\pi}{k} \, Y \mathcal{P}^\prime - Y^{\prime \prime \prime}   \right)\, u\,, \\
	\chi & =&   \frac{1}{\sqrt{2}} \, \left(-\frac{2\pi}{k}Y\mathcal{P}+Y+Y''\right)\,, \hspace{1cm}
	\varepsilon^1 =  -\frac{\ell}{\sqrt{2}} \, \left(\frac{2\pi}{k}Y\mathcal{P}+Y-Y''\right)\,, \\
	\n \chi^1 &=& \frac{1}{\ell} T' \,, \hspace{4.7cm} \varepsilon^2 =  -\ell \, Y' \,, \\
	\n   \chi^2  &=& -\frac{1}{\sqrt{2}\, \ell} \, \left(\frac{2\pi}{k}\left(Y \mathcal{J}+T\mathcal{P}\right)+T-T''\right) -  \frac{1}{\sqrt{2}\, \ell}\, \left( Y^\prime\left( \frac{2\pi}{k} \mathcal P + 1  \right) +\frac{2\pi}{k} \, Y \mathcal{P}^\prime - Y^{\prime \prime \prime}   \right)\, u\,, 
	\ea
	where we defined $\mathcal P \to \frac{4\pi}{k}\, \mathcal P$ and $\mathcal{J} \to  \frac{4\pi}{k}\mathcal J $ for mere aesthetic convenience. Notice that the set of equations is reduced to having only two independent functions $Y=Y(\phi)$ and $T=T(\phi)$, which, in virtue of the isomorphism, are associated with the infinite-dimensional extensions of the Poincaré gravity: superrotations and supertranslations. The functions $\mathcal J$ and $\mathcal P$ must transform under the gauge transformation \eqref{eq:gaugetransf} as
	\bseq\label{eq:transformationslawAdSC}
	\ba
	\delta_\lambda \mathcal{P} &= & \mathcal{P}^\prime Y + 2 \mathcal{P} Y^\prime - \frac{k}{2\pi}Y^{\prime \prime \prime}\,,\\
	\delta_\lambda \mathcal J &=& \mathcal{P}^\prime T +2\mathcal P T^\prime +\mathcal{J}^\prime Y + 2\mathcal J Y^\prime- \frac{k}{2\pi}T^{\prime \prime \prime}\,.
	\ea
	\eseq
	These symmetry transformations correspond to those of the $\mathfrak{bms}_3$ algebra \cite{Ashtekar:1996cd, Barnich:2006av}. It is worth mentioning that the presence of the AdS Carroll radius $\ell$ has disappeared completely from these transformation laws. This is aligned with the asymptotic structure of the Poincaré symmetry where does not appear the cosmological constant. This result verifies that the isomorphism of \eqref{eq:mapgenerators} also works for studying the asymptotic structure of the AdS Carroll spacetimes.\\
	In the gauge \eqref{eq:gaugetransformation}, the only remaining field equation to solve is $f_{u\phi}  =0$. This gives the time evolution for $\dot{a}_\phi$, which can be interpreted as a gauge transformation for $a_\phi$ parametrized by the Lagrange multiplier $a_u$ \cite{Bunster:2014mua}. This implies that the asymptotic form of $a_\phi$ will be maintained along different time slices and with $a_u$ of the form
	\be
	a_u = \lambda[\mu_{\mathcal P}\,, \mu_{\mathcal J}]\,,
	\ee 
	with $\lambda[\mu_{\mathcal P}\,, \mu_{\mathcal J}]$ corresponding to the solution \eqref{eq:transformationslawAdSC} by doing the change $T \to \mu_{\mathcal P}$ and $Y \to \mu_{\mathcal J}$, where $\mu_{\mathcal P}\,, \mu_{\mathcal J}$ stand for arbitrary functions of $\mathcal P$ and $\mathcal J$ and their derivatives, and that are assumed to be fixed at the boundary. Then, the consistency of the field equation $f_{u\phi}  =0$ implies that
	\bseq\label{eq:consistencyeqs}
	\ba
	\dot{\mathcal P} &= & 2\mathcal P \mu_{\mathcal J}^\prime+ \mathcal{P}^\prime \mu_{\mathcal J}- \frac{k}{2\pi}\, \mu_{\mathcal J}^{\prime \prime \prime}\,,\\
	\dot{\mathcal J} &= & 2\mathcal J \mu_{\mathcal J}^\prime+ \mathcal{J}^\prime \mu_{\mathcal J}+2\mathcal P \mu_{\mathcal P}^\prime+ \mathcal{P}^\prime \mu_{\mathcal P} - \frac{k}{2\pi}\, \mu_{\mathcal P}^{\prime \prime \prime}\,.
	\ea
	\eseq
	Due to the $r$-independence on these equations, they are usually referred to as the asymptotic field equations of motion. The parameters of the asymptotic symmetries must hold the following conditions
	\bseq\label{eq:consistencyparameters}
	\ba
	\dot Y &=& \mu_{\mathcal J} Y^\prime - \mu_{\mathcal J}^\prime Y\,,\\
	\dot T &=& \mu_{\mathcal J} T^\prime - \mu_{\mathcal J} T + \mu_{\mathcal P} Y^\prime -\mu_{\mathcal P}^\prime Y\,.
	\ea
	\eseq
	To sum up, like the relativistic case, the set of equations \eqref{eq:consistencyeqs} provide a consistency condition for the Poisson structure of the system, with $Y $ and $ T $ being dependent on the dynamical fields and their derivatives.

	\subsection{Canonical charges}
	Now, we want to compute the variation of the canonical charges associated with the gauge transformations (\ref{eq:transformationslawAdSC}). With the determination of the gauge parameters \eqref{eq:gaugetransf}, leaving invariant the boundary condition \eqref{eq:bbccAdSCarroll}, we derive the canonical charges in AdS Carroll gravity by following the standard Regge-Teitelboim prescription~\cite{Regge:1974zd,Banados:1994tn,Bergshoeff:2019rdb, Frodden:2019ylc}
	\be
	\delta Q[\Xi] =- \frac{k}{2\pi} \, \oint \mathrm{d}\phi \, \Big\langle \Xi\,,  \delta A_\phi \Big\rangle\,,
	\ee
	where $ \langle \,,\rangle $ is the bilinear form (\ref{Eq:AdSpairing}). Applying the gauge transformation \eqref{eq:gaugetransformation} on this last equation, we introduce the connection \eqref{eq:AdSCarrollbdrtyconditions} and \eqref{eq:gaugeparameterwithoutr}, then we get 
	\be
	\delta Q[\lambda] =- \frac{k}{2\pi} \, \oint \mathrm{d}\phi \, \Big\langle \lambda,  \delta a_\phi \Big\rangle\,.
	\ee
	Using the boundary conditions \eqref{eq:AdSCarrollbdrtyconditions} and the gauge parameters \eqref{eq:gaugetransf}, then the charge reduces to the following simple expression
	\be
	\delta Q =  -\oint \mathrm{d}\phi \, \left(T\, \delta \mathcal P + Y \, \delta \mathcal J \right)\,.
	\ee
	Since the functions $T$ and $Y$ do not depend on the fields, this expression can easily be integrated into phase space and leads to the following finite canonical charge
	\be\label{eq:chargeAdSC}
	Q =  -\oint \mathrm{d}\phi \, \left(T\, \mathcal P + Y \,  \mathcal J \right)\,.
	\ee
	It is trivial to check that this is conserved in retarded time, namely $\dot{Q}[\lambda]=0$, where we have used the equations motion \eqref{eq:consistencyeqs}
	and the consistency conditions for the symmetry parameters \eqref{eq:consistencyparameters}. 
	\subsection{Asymptotic symmetry algebra}
	The algebra of the conserved charges \eqref{eq:chargeAdSC} can be directly obtained from their corresponding Poisson brackets via the shortcut equation
	\be
	\{ Q[\lambda_1]\,, Q[\lambda_2]\} = \delta_{\lambda_2}\, Q[\lambda_1]\,.
	\ee
	With the help of \eqref{eq:transformationslawAdSC} and the charge \eqref{eq:chargeAdSC}, the boundary conditions \eqref{eq:AdSCarrollbdrtyconditions} yields to the following non-vanishing Poisson brackets
	\bseq
	\ba
	\{ \mathcal{J}(\phi) \,, \mathcal{J}(\bar \phi) \} &=& -2\mathcal{J}(\phi)\, \partial_\phi \left(\phi - \bar \phi \right) - \delta \left(\phi -\bar \phi \right)\, \partial_\phi \mathcal J (\phi)\,,\\
	\{ \mathcal{J}(\phi) \,, \mathcal{P}(\bar \phi) \} &=&-2 \mathcal{P}(\phi)\, \partial_\phi \delta \left(\phi - \bar \phi \right) - \delta \left(\phi - \bar \phi \right)\, \partial_\phi \mathcal{P} (\phi) + \partial^{3}_{\phi} \, \delta \left(\phi -\bar \phi \right)\,.
	\ea
	\eseq
	Expanding the fields, delta distribution and their derivatives in terms of Fourier modes as
	\ba
	\mathcal{J}(\phi) &=& \frac{1}{2\pi} \sum_{n\in \mathbb{Z}} \mathcal{J}_ne^{-in\phi} \,, \hspace{0.5cm} \mathcal{P}(\phi) = \frac{1}{2\pi} \sum_{n\in \mathbb{Z}} \mathcal{P}_ne^{-in\phi}\,,\\
	\delta (\phi - \bar \phi) &=& \frac{1}{2\pi} \sum_{n\in \mathbb{Z}} e^{-in(\phi-\bar \phi)}\,,
	\ea
	we get the only non-vanishing commutators
	\ba\label{eq:BMS3algebra}
	i\,\{\mathcal J_m\,, \mathcal J_n \} &=& (m-n)\, \mathcal J_{m+n}\\
	i\, \{\mathcal J_m\,, \mathcal P_n \} &=& (m-n)\, \mathcal P_{m+n} +k\, m^3 \delta_{m+n,0}\,,
	\ea
	which is the $\mathfrak{bms}_3$ algebra \cite{Barnich:2010eb, Barnich:2006av,Ashtekar:1996cd}. This corresponds to the asymptotic symmetry algebra of AdS Carroll spacetimes in $(2+1)$-dimensions for the particular boundary conditions \eqref{eq:AdSCarrollbdrtyconditions}.\footnote{See also an asymptotic analysis for general relativity with a cosmological constant in \cite{Perez:2022jpr,Fuentealba:2022gdx}.}  Note the non-presence of the cosmological constant. This is in agreement with the fact that the massless AdS Carroll particle possesses the same infinite-dimensional symmetry as the massless Carroll particle: the so-called conformal Carroll particle in \cite{Casalbuoni:2023bbh, Bergshoeff:2014jla}. Then, in virtue of the isomorphism presented in \cite{Duval:2014uva} the symmetries of the conformal Carroll particle contain the BMS$_3$ symmetry \cite{Duval:2014uva}.
	
	Since Poincaré algebra is contained as a finite sub-algebra in $\mathfrak{bms}_3$ algebra, then due to the isomorphism \eqref{eq:mapgenerators} is expected that AdS Carroll algebra does as well. The generators of AdS Carroll 
	are recovered as 
	\bseq
	\ba
	H&=& \frac{1}{\sqrt{2}}\, \mathcal P_0 - \frac{1}{2\sqrt{2}} \, \left(\mathcal P_1+\mathcal P_{-1} \right)\\
	P_1 &=& - \frac{1}{\sqrt{2}\, \ell }\, \mathcal J_0 - \frac{1}{2\ell \, \sqrt{2}} \, \left(\mathcal J_1+\mathcal J_{-1} \right)\\
	P_2 &= & -\frac{1}{2\, \ell}\left(\mathcal J_1-\mathcal J_{-1} \right)\\
	J&=& -\frac{1}{\sqrt{2}}\, \mathcal J_0 + \frac{1}{2\sqrt{2}} \, \left(\mathcal J_1+\mathcal J_{-1} \right)\\
	B_1 &=& \frac{\ell}{2} \, \left(\mathcal P_1-\mathcal P_{-1} \right)\\
	B_2 &= &-\frac{\ell}{2} \, \mathcal P_0 - \frac{\ell}{2\sqrt{2}} \, \left(\mathcal P_1+\mathcal P_{-1} \right)\,.
	\ea
	\eseq

	%
	\section{AdS Carroll geometry structure}\label{Geometrysection}
	The AdS Carroll boundary conditions derived from the map with the help of the Chern-Simons formulation can be translated into the metric formulation by extracting the dreibein $e_{\mu}^{a}$ from the connection $A$. Like the gauge formulation for Einstein gravity, having introduced locally the one-form gauge connection $A$ valued on the AdS Carroll algebra, now one should identify some of the generators with a basis of the tangent space of the manifold. Here the one forms $\{\tau_\mu \,, {e_{\mu}}^{a} \}$ constitute the basis of the cotangent space. Then, the dual basis of the tangent space is denoted by $\{\tau^\mu \,, {e^\mu}_{a} \}$ obeying the orthogonality and completeness conditions
	\ba
	&& {e_\mu}^{a} \, {e^\mu}_{b}  = \delta^a_b\,, \hspace{0.5cm} \tau_\mu \, \tau^\mu =1\,, \hspace{0.5cm} \tau^{\mu}\, {e_\mu}^{a} =0\,, \hspace{0.5cm} \tau_\mu \, {e^\mu}_{a} = 0\,,\\
	&& {e_\mu}^{a} \, {e^\nu}_{a} + \tau_\mu \, \tau^\nu = \delta^\nu_\mu\,.
	\ea
	Here the local-frame $\{ \tau_\mu\,, {e_\mu}^{a}\}$ and the dual-local frame $\{ \tau^\mu\,, {e^\mu}_{a}\}$ allows us to map spacetime indices into tangent spacial ones. The degenerate spatial Carroll metric $h_{\mu \nu}$ is given by
	\be\label{eq:carrollmetric}
	h_{\mu \nu}  = {e_\mu}^{a} \, {e_\nu}^{b} \, \delta_{ab} \,,  
	\ee
	with a null vector $\tau^\mu$ fulfilling
	\be
	h_{\mu \nu} \, \tau^\nu =0\,.
	\ee
	To differentiate with the Bondi-like coordinates, we introduce the Boyer-Lindquist coordinates denoted by $x^\mu =(t\,,r\,,\varphi )$, with $-\infty < t <\infty$, $r\geq 0$, and the cyclic coordinate $\varphi \sim \varphi +2\pi$. In these coordinates, the flat cosmology metric is given by \cite{Hoseinzadeh:2014bla,Concha:2018zeb}
	\be\label{eq:Poincarecosmologysolution}
	ds^2 = -N^2 dt^2 + \frac{dr^2}{N^2} + r^2 \left(d\varphi +N_\varphi\, dt \right)^2
	\ee
	where 
	\be
	N^2 = -M+\frac{J^2}{4r^2} \,, ~~~~~N_\varphi = -\frac{J}{2r^2}\,,
	\ee
	with $M$ and $J$ being integration constants, and the spacetime boundary is located at $r\to \infty$ limit. The dreibein is chosen to be
	\ba
	\tilde{e}^0 &=&N\, \mathrm{d}t\\
	\tilde{e}^1 &=&N^{-1}\, \mathrm{d}t\\
	\tilde{e}^2  &=&r\,\left(\mathrm{d}\varphi+ N_\varphi \mathrm{d}t  \right)\,,
	\ea
	and the torsion-free spin connection
	\ba
	\tilde{\omega}^{0} &=& N\, \mathrm{d}\varphi \\
	\tilde{\omega}^{1} &=& \frac{J}{2r^2 N}\, \mathrm{d}r \\
	\tilde{\omega}^{2} &=&- \frac{J}{2r}\, \mathrm{d}\varphi \,.
	\ea
	Utilizing the map \eqref{mapCP}, the AdS Carroll gauge fields are written in terms of the solution as
	\bseq
	\ba
	\tau  &=&N\, \mathrm{d}t\\
	e^1 &=& - \frac{\ell\, J}{2r}\, \mathrm{d}\varphi\label{eq:e1BL} \\
	e^2 &=&    \frac{\ell\, J}{2r^2 N}\, \mathrm{d}r \label{eq:e2BL} \\
	\omega &=& N\, \mathrm{d}\varphi \\
	\omega^1 &= & -\frac{1}{\ell \,N }\, \mathrm{d}t\\
	\omega^2 & =& -\frac{1}{\ell} \,r\,\left(\mathrm{d}\varphi+ N_\varphi \mathrm{d}t  \right)\,,
	\ea
	\eseq
	Considering the Eqs.~\eqref{eq:e1BL} and \eqref{eq:e2BL}, we construct the degenerate spatial AdS Carroll metric given by
	\be\label{eq:AdSmetric}
	h_{\mu \nu} ={e_\mu}^a {e_\nu}^b \delta_{ab}= (e^1)^2 + (e^2)^2=0 \times \mathrm{d}t^2 +\frac{dR^2}{\left(-M+ \frac{R^2}{\ell^2}\right)}  +R^2\, \mathrm{d}\varphi^2\,.
	\ee
	where we defined $R= J\ell/2r$. The metric $h_{\mu \nu}$ is degenerate and has the null vector 
	\be
	\tau^\mu = \left(\frac{1}{N},0,0\right)\,.
	\ee
	We see that this metric has the same form as the transverse part of the BTZ black hole metric. It could correspond with the solution of the magnetic gravity in $(3+1)$-dimensions \cite{Perez:2022jpr}, which admits as a solution the transverse part of the Schwarzschild black hole was found. On the other hand, the Carroll structure associated with the boundary conditions \eqref{eq:AdSCarrollbdrtyconditions} can be obtained via the map \eqref{eq:mappinglightcone}. Then, the gauge fields for AdS Carroll algebra in terms of the solution \eqref{eq:solutioninBMSgauge}, yielding  
	\ba
	\tau &= & \frac{1}{\sqrt{2}}\, \left(-\mathrm{d}r + \left(\frac{2\pi}{k}\,\mathcal{P} -1\right) \, \mathrm{d}u + \frac{2\pi}{k}\, \mathcal{J} \, \mathrm{d}\phi\right)\\
	e^1 &= & -\frac{\ell}{\sqrt{2}}\, \left(\frac{2\pi}{k}\,\mathcal{P} +1\right)\, \mathrm{d}\phi   \\
	e^2 &= & 0 \\ 
	\omega  &= & \frac{1}{\sqrt{2}} \left(1-\frac{2\pi}{k}\,\mathcal{P} \right) \, \mathrm{d}\phi\\
	\omega^1 &=& -\frac{1}{\ell}  \, r \, \mathrm{d}\phi\\
	\omega^2& =&-  \frac{1}{\sqrt{2}\, \ell} \, \left(-\mathrm{d}r + \left(\frac{2\pi}{k}\,\mathcal{P} +1\right) \, \mathrm{d}u + \frac{2\pi}{k}\,\mathcal J \, \mathrm{d}\phi\right)\,.
	\ea
	
	The degenerate metric for this solution now becomes
	\be
	h_{\hat{\mu} \hat{\nu}} = {e^a}_{\hat{\mu}}  \, {e^b}_{\hat{\nu}}\, \delta_{ab} = 0\times \mathrm{d}u^2 +0\times \mathrm{d}u \mathrm{d}r +\frac{\ell^2}{2}\, \left(\frac{2\pi}{k} \mathcal{P} + 1\right)^2\, \mathrm{d}\phi^2
	\ee
	with a null vector given by
	\be
	\tau^{\hat{\mu}} = -\sqrt{2}\left(1\,, \frac{2\pi}{k } \mathcal P\,,0\right)\,,
	\ee
	such that $h_{\hat{\mu}  \hat{\nu}} \, \tau^{\hat{\nu}} =0$. Unlike the relativistic flat cosmology case, the rank of this metric is one, but the rank of \eqref{eq:AdSmetric} is two, so there is no change of coordinates relating to both metrics.

	\section{Discussion}\label{Section:discussion}

	In this paper, we utilize the isomorphism between AdS Carroll and Poincaré algebras to map the \((2+1)\)-dimensional flat cosmology into AdS Carroll geometry. By considering the Chern-Simons formulation of \((2+1)\)-dimensional Poincaré gravity, the isomorphism among generators induces a mapping between the gauge fields, resulting in an exchange between the spatial dreibein and the spin connection. This suggests a set of boundary conditions for the AdS Carroll gravity theory.
	
	The asymptotic structure of the resulting AdS Carroll spacetime is non-trivial, with the symmetry algebra being the $\mathfrak{bms}_3$ algebra, independent of the cosmological constant. These results indicate that the isomorphism between Poincaré and AdS Carroll algebras can be extended to infinite-dimensional asymptotic algebras. This aligns with the observation that the massless AdS Carroll particle possesses the same infinite-dimensional symmetry as the massless Carroll particle, referred to as the conformal Carroll particle in \cite{Bergshoeff:2014jla, Bergshoeff:2015wma, Casalbuoni:2023bbh}. Furthermore, according to the isomorphism presented in \cite{Duval:2014uva}, the symmetries of the conformal Carroll particle include the BMS\(_3\) symmetry. Finally, we construct the Carroll structure of the resulting solution in the AdS Carroll gravity using two different sets of coordinates.
	
	It would be interesting to further analyze the physical aspects of Carroll spacetimes. As emphasized previously, the physical interpretations of the homogeneous spaces associated with the AdS Carroll and Poincaré algebras differ significantly. Therefore, determining the notions of singularities and thermodynamic properties in AdS Carroll gravity is a natural extension of this work. For instance, determining holonomy conditions for these spacetimes could provide insights into their topology, as discussed in \cite{Briceno:2024ddc}.
	
	Another avenue worth exploring is studying the map in \((1+1)\)- or higher-dimensional spacetimes. In particular, applying this isomorphism to any \((1+1)\)-dimensional BF theory and interpreting its corresponding solutions could be of interest. Additionally, the supersymmetric extension of this isomorphism could prove to be valuable.

	\section*{Acknowledgements}
	We are grateful to Carlos Batlle, Jos\'e Figueroa-O'Farrill, Oscar Fuentealba, and Pablo Rodríguez for enlightening discussions and comments. JG is grateful to the Erwin Schrödinger Institute in Vienna for the invitation to the workshop Carrollian Physics and Holography, where this project has been initiated. DH thanks the Departament de F\'isica Qu\`{a}ntica i Astro\'isica of Universitat de Barcelona for hospitality during the completion of this work. The research of JG was supported in part by PID2022-136224NB-C21 and by the State Agency for Research of the Spanish Ministry of Science and Innovation through the Unit of Excellence Maria de Maeztu 2020-2023 award to the Institute of Cosmos Sciences (CEX2019- 000918-M). DH is supported by the Icelandic Research Fund Grant 228952-053 and by the University of Iceland Research Fund. L.A. is supported by Fondecyt and SIA grants N $^{\mathrm{o}}$3220805 and N$^{\mathrm{o}}$85220027, respectively.

	\appendix

	\section{Poincaré algebra in lightcone form}\label{section:appendixlightcone}
	The $(2+1)$-Poincaré algebra spanned by the Lorentz $\tilde{J}_A$ and rotational generators $\tilde{P}_A$, with $A=0,1,2$, is given by \eqref{eq:Poincarealgebranormal}. It can be described in a lightcone form. Let us denote lightcone generators with a hat symbol: the Lorentz generators $\hat{J}_A$ and rotational generators $\hat{P}_A$ which satisfy the following non-vanishing commutation relations given by
	\be\label{eq:Poincarélightcone}
	\left[\hat{J}_A\,, \hat{J}_B\right] = \epsilon_{ABC}\, \hat{J}^C\,, \hspace{1.3cm} \left[\hat{J}_A\,, \hat{P}_B\right] = \epsilon_{ABC} \hat{P}^C\,,
	\ee
	along with an invariant and non-degenerate bilinear form 
	\be
	\langle \hat{J}_A \,,  \hat{J}_B\rangle =- \alpha_0 \, \hat{\eta}_{AB}\,, \hspace{1cm}  \langle \hat{J}_A \,,  \hat{P}_B\rangle =- \alpha_1 \, \hat{\eta}_{AB}\,, 
	\ee
	As we see the structure of the algebra and bilinear form are the same as the standard Poincaré, but to raise and lower indexes we use the off-diagonal Minkowski metric $\hat{\eta}_{AB}$, given by
	\be
	\hat{\eta}_{AB} =\left( \begin{matrix}
		0 & 1 & 0\\
		1 & 0 & 0\\
		0& 0 & 1
	\end{matrix} \right)\,.
	\ee
	For instance $\hat{P}_0 = \hat{\eta}_{01} \, \hat{P}^{1} = \hat{P}^1$. The relationship between lightcone Poincaré generators $\{\hat{J}_{A} \,, \hat{P}_{A}\} $ and standard Poincaré generators $\{\tilde{J}_A\,, \tilde{P}_A \}$ is given by
	\bseq\label{eq:mapPoincareandlightcone}
	
	\ba
	\hat{J}_0 &= & -\frac{1}{\sqrt{2}}\, \left(\tilde{J}_0 + \tilde{J}_2 \right) \,, \hspace{2cm} \hat{P}_0 =  \frac{1}{\sqrt{2}}\, \left(\tilde{P}_0 + \tilde{P}_2 \right)   \\
	\hat{J}_1 &= & \frac{1}{\sqrt{2}}\, \left(\tilde{J}_0 - \tilde{J}_2 \right)\,, \hspace{2.4cm} \hat{P}_1 = - \frac{1}{\sqrt{2}}\, \left(\tilde{P}_0 - \tilde{P}_2 \right)        \\
	\hat{J}_2 &= & -\tilde{J}_1 \,, \hspace{4cm} \hat{P}_2 =  \tilde{P}_1
	\ea
	\eseq
	whose inverse relationship is given by
	\bseq\label{eq:inversemapPoincaréandlightcone}
	\ba
	\tilde{J}_0 &= &- \frac{1}{\sqrt{2}}\, \left(\hat{J}_0 - \hat{J}_1 \right) \,, \hspace{2.4cm} \tilde{P}_0 =  \frac{1}{\sqrt{2}}\, \left(\hat{P}_0 - \hat{P}_1 \right)   \\
	\tilde{J}_1 &= & -\hat{J}_{2} \,, \hspace{4.4cm} \tilde{P}_1 = \hat{P}_{2}      \\
	\tilde{J}_2 &= & - \frac{1}{\sqrt{2}}\, \left(\hat{J}_0 + \hat{J}_1 \right)\,, \hspace{2.4cm} \tilde{P}_2 = \frac{1}{\sqrt{2}}\, \left(\hat{P}_0 + \hat{P}_1 \right)  
	\ea
	\eseq

\end{document}